# LSST optical beam simulator

J. A. Tyson[a], J. Sasian[b], K. Gilmore[c], A. Bradshaw[a], C. Claver[d], M. Klint[a], G. Muller[d], G. Poczulp[f], E. Resseguie[a]
[a]UC Davis, [b]UA Optical Sciences Center, [c]SLAC, [d]LSST, [f]NOAO

## ABSTRACT

We describe a camera beam simulator for the LSST which is capable of illuminating a 60mm field at f/1.2 with realistic astronomical scenes, enabling studies of CCD astrometric and photometric performance. The goal is to fully simulate LSST observing, in order to characterize charge transport and other features in the thick fully-depleted CCDs and to probe low level systematics under realistic conditions. The automated system simulates the centrally obscured LSST beam and sky scenes, including the spectral shape of the night sky. The doubly telecentric design uses a nearly unit magnification design consisting of a spherical mirror, three BK7 lenses, and one beam-splitter window. To achieve the relatively large field the beam-splitter window is used twice. The motivation for this LSST beam test facility was driven by the need to fully characterize a new generation of thick fully-depleted CCDs, and assess their suitability for the broad range of science which is planned for LSST. Due to the fast beam illumination and the thick silicon design [each pixel is 10 microns wide and over 100 microns deep] at long wavelengths there can be effects of photon transport and charge transport in the high purity silicon. The focal surface covers a field more than sufficient for a 40x40mm LSST CCD. Delivered optical quality meets design goals, with 50% energy within a 5 micron circle. The tests of CCD performance are briefly described.

## 1. INTRODUCTION

The Large Synoptic Survey Telescope (LSST) three mirror optics creates an f/1.2 centrally obscured beam covering a 9.6 square degree field of view. The half cone angle of the beam spans 14.2 to 23.6 degrees. The focal plane consists of 189 novel 4Kx4K format 100 micron thick CCDs with 10 micron pixels on fully depleted high resistivity silicon. Each CCD is segmented into 16 sub-CCDs with their own readout, enabling a 2 second read time for the entire 3.2 Gpixel camera. We must fully characterize the optoelectronic performance of the LSST CCDs prior to camera and data pipeline construction. In order to optimize science data analysis algorithms for weak gravitational lens galaxy shear, precision understanding of the PSF shape transfer function is also required.

In this report we describe an LSST camera beam simulator on an optical bench which is capable of illuminating full CCDs with realistic astronomical scenes, and which enables studies of CCD astrometric and photometric performance. It is always important to test new imaging detectors for a mosaic camera before device acceptance and constructing the mosaic. This is particularly true of the LSST CCDs due to the fast beam illumination and the thick silicon design: each pixel is 10 microns wide and over 100 microns deep. Photoelectron transport anomalies in thick chips produce a mapping between input optical and output electrical coordinates. At long wavelengths there can be effects of photon transport in addition to photoelectron transport in the high purity silicon because of the long absorption length for photons near the band gap. In order to study science capability, realistic sky scenes and spectral distributions need to be projected onto the CCD focal plane. Wavelengths of 0.7 – 0.9 micron are important for precision weak gravitational lens measurements and require precision control of detector PSF shape systematics. For this we have designed and built a wide-field f/1.2 re-imaging system delivering an optical beam identical to that of the LSST camera. The fully automated system simulates the entire LSST operation, including artificial galaxies and stars with approximately black-body spectra superimposed on a spatially diffuse night sky emission with its complex spectral features.

The specifications of the LSST beam simulator are to simulate the beam of the LSST telescope and camera optics, including the obscuration, and to provide a nearly diffraction limited image over a 60 mm diameter field covering an entire 4Kx4K ten micron pixel chip. The system must be capable of projecting spots which are a small fraction of the 10 micron pixel over several wave-bands. The desired wavelength range for exploring CCD performance related to weak gravitational lens science applications is 0.6 to 1.2 micron, but not simultaneously (allowing refocus in between band changes). In addition, the system must be compact enabling installation on a laboratory optical table.

The beam simulator images a sky scene source which is up to 2m away onto the CCD inside a dewar. All system parameters are tunable over the entire range of LSST operations, including CCD temperature, parallel readout of all segments, readout rate, clock timing, multiple image dithering, filter passbands, sky spectra, etc.

Machining and assembly of the LSST Simulator was done by the NOAO machine and optics shops. Subsequent analysis of CCD images of 5 micron pinholes at various field points was performed using a 40x microscope with commercial 4 µm pixel CCD camera to reimage the Petzval plane. These analyses have confirmed that the simulator's performance conforms to the specifications regarding fraction of enclosed energy and aberration correction, particularly coma and linear astigmatism.

It was possible to achieve the specifications by utilizing a nearly unit magnification system consisting of a spherical mirror, three BK7 lenses, and one beam-splitter window. To achieve the relatively large field the beam-splitter window is used twice and thus the system loses about 80% of the light but this is compensated by the use of brighter sources. Compensation is achieved by moving the first lens and reticle relative to the rest of the system.

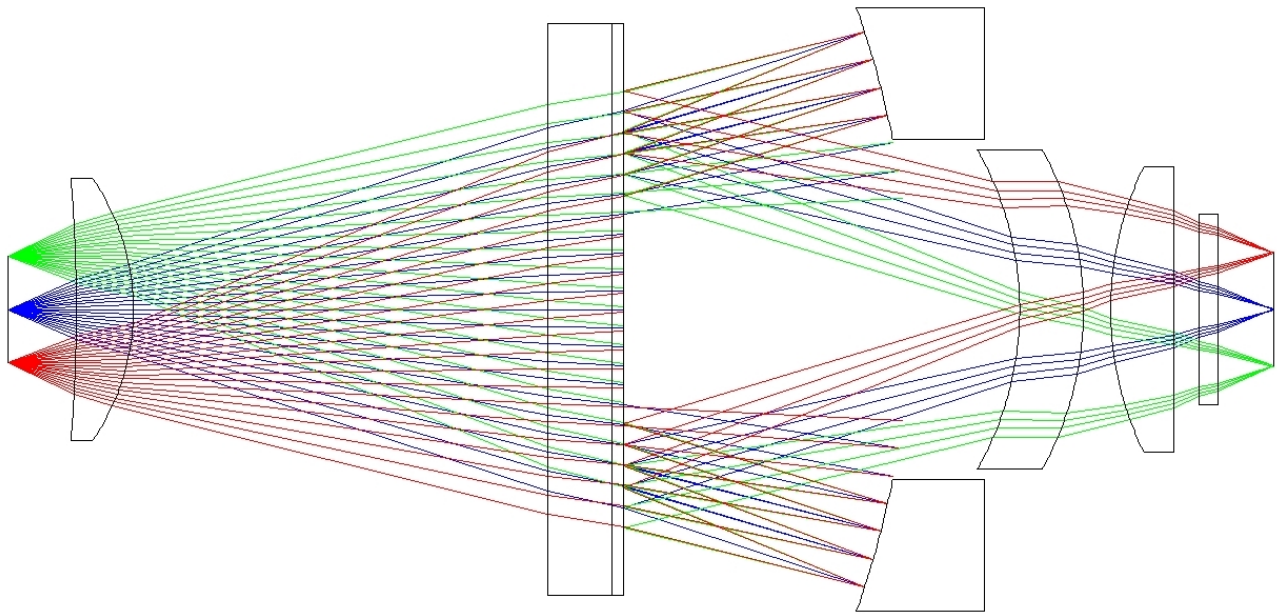

Figure 1. Rays in the double telecentric f/1.2 reimager, from source reticle mask (left) to focal plane (right).

A variety of optical designs were considered and the final design is shown in Figure 1. Light from the object on the left side, here a reticle mask with pinholes, passes the first lens and then 60% of it is transmitted by the beam-splitter. Light then reaches the mirror and is reflected back towards the beam-splitter where about 40% is reflected toward the lenses shown on the right side. After light passes through these lenses it enters the cryogenic dewar (not shown here) passing through the quartz dewar window to a focal plane. All lenses and the dewar window are AR coated and the mirror is aluminum coated. A light baffle in the center of the beam-splitter blocks direct light from reaching the image plane; it also simulates the central obscuration by the secondary mirror of the LSST telescope.

The idea of exploiting a spherical mirror and beam-splitter as a unit magnification, high numerical aperture, re-imager can be traced to Burch[1]. Dyson[2] extended this by introducing further lenses to flatten the field and correct astigmatism.

Here both approaches are combined to provide a fast beam over a large field. Other 'lossless' approaches to the problem can become physically large and require many more optical elements and aspheric surfaces.

Analysis of stray light indicates that no significant problems should be expected if proper baffling is integrated into the system. The different wave-bands are accommodated by shifting the first lens and refocusing the reticle mask carrying the array of pinhole stars or a scene with gray scale galaxies and stars. The optical length of the system is 665.7 mm and the maximum diameter of the optics is 317.5 mm.

The system is aligned to the window in tilt since angular misalignment will introduce coma. This is a useful degree of freedom to remove any coma residual in the system. At 0.1 degree of dewar window tilt there is no appreciable change in system performance. The alignment mechanism provides an angular resolution alignment of better than 0.1 degree in x and y. This must be provided since there can be a small uniform coma residual in the system that can be removed by tilting the system with respect to the dewar window. However, we have not found any measureable coma at normal incidence.

While it is possible to get the CCD chip mechanics approximately aligned inside the dewar, there will be tilt errors at the 20 micron/40mm (0.03 deg) level that must be corrected by tilting the dewar once it is cold. We require a way of measuring and adjusting the plane of the chip/carrier at room temperature with the dewar open. We want to simulate CCD surface tilt at the level of the camera specs (less than 10 microns / 40mm peak to valley). The dewar manual tilt adjust must therefore be 2-axis and cover this range of angles.

## 2. IMAGE QUALITY

Figure 2 shows the ZEMAX calculated optical performance in the R band (680-820 nm) for a point source. The black line is the diffraction limit, and the three color lines show the performance at various field positions in the focal plane: on-axis (blue), 12mm radius (green), 24mm radius (red), and 28mm radius (yellow). Uniform high quality images are delivered over the entire chip in the focal plane.

Figure 3 shows the observed fraction of enclosed energy for 5 μm pinholes at 0, 10, 20 and 28 mm radii for the R band (680 – 820 nm). As stated, all field points show half the energy enclosed within a 5 μm diameter.

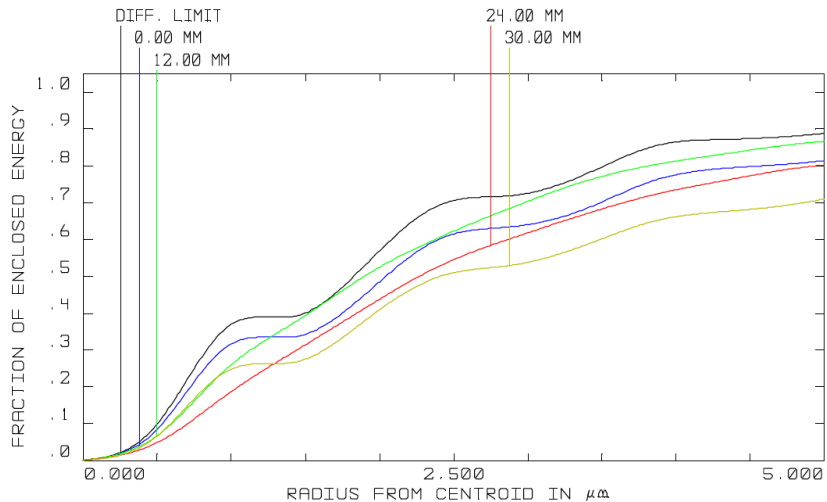

Figure 2. ZEMAX calculated performance for a point source in the reticle mask as a function of field position. The top curve is the diffraction limit.

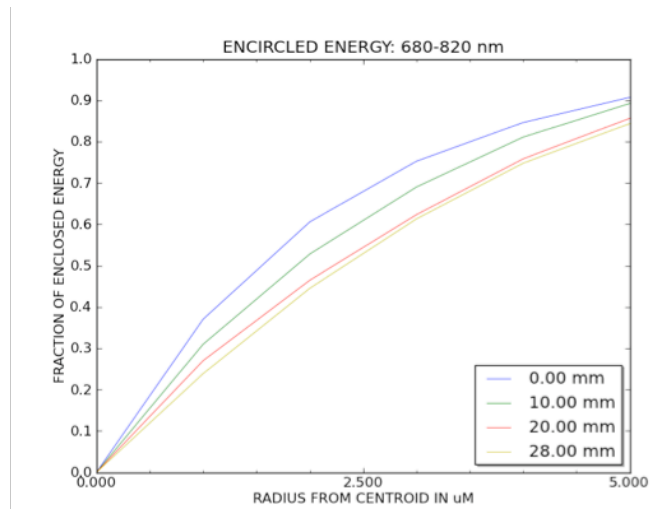

Figure 3. Observed fraction of enclosed energy for a 5 μm pinhole in the reticle mask.

Figure 4 (below) shows the result of a magnified examination of the output beam at focus with a microscope and camera with 0.1 micron resolution.

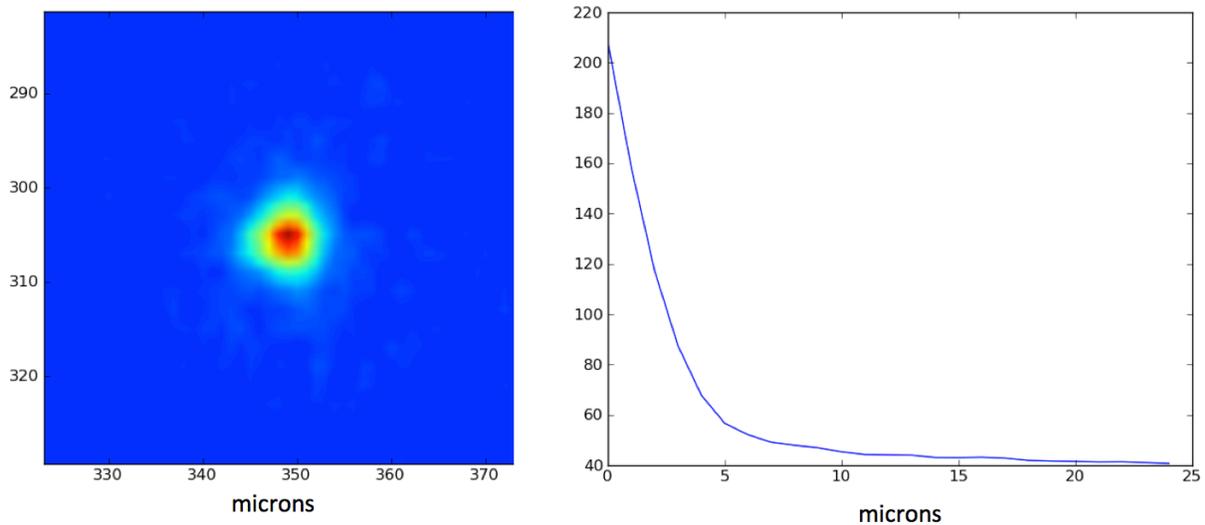

Figure 4. For a ~5 μm hole, half width at half maximum of the average profile is 1.86 microns. This was done using a x100 microscope with commercial 4 μm pixel CCD camera to reimage the Petzval plane. The figure on the left shows the focused image, while the plot on the right shows the intensity vs radius.

Since the system is telecentric, light can be reflected back by the sensor all the way to the spherical mirror and will return almost on its own path to form a faint ghost image near the primary image. However these ghosts are far out of focus, and the spot diagrams of ghost images are larger than 2 mm in diameter. The ghost light is spread over a much larger area than the in-focus PSF by the ratio 3 sq.mm / 20 sq. microns = 1.5e5. This surface brightness ratio of 7e-6 is further reduced by the reflectivity of the AR coated CCD. So the average surface brightness of the ghost is many orders of magnitude lower than the main in-focus image. Reflection on the reticle mask can potentially give a focused ghost image. This is an inherent problem in doubly telecentric systems. The beam splitter window helps to reduce the problem. A large reduction is obtained via a black non-reflecting coating on the inside surface of the reticle mask.

Thus the quality of these expected PSFs is more than sufficient to meet the needs for LSST CCD testing. First, they are totally dominated by diffraction, so any improvements that can be made would be very small. Second, if one convolves these with a 2-5 micron spot (size of the pin hole as the source) the minimal asymmetric effects seen off axis are that much more diluted. The PSF errors are dominated by the sensor charge diffusion, as intended.

## 3. DIFFUSE NIGHT SKY SPECTRAL SIMULATION

In addition to the stars and galaxies (with their black body spectrum provided by the scattering sphere illumination of the reticle mask), there is a need to simulate the diffuse night sky and its complex spectrum. This is done via a scanning monochromator with a 640-1200nm range with 0.1nm spectral resolution. The complex sky emission spectrum, dominated by multiple vibration-rotation bands, as well as the LSST filter (*rizy*) transmission curve, is simulated during an exposure. Constant illumination over the focal plane is achieved by injecting "sky" light in the pupil plane. Illumination emanating from the periphery of the pupil plane mirror directed towards the splitter mirror is spread uniformly in the focal plane (Fourier transform of the pupil plane). Four optical fibers arranged around the outside edge of the mirror, aimed upstream towards the back side of the beam splitter mirror will produce the uniform (within 0.5%) illumination across 40x40 mm of the focal plane as seen in Figure 5. Figure 6 shows the orientation of the four fibers around the edge of the primary mirror. Using four fibers, measurements of the delivered flat field illumination is in good agreement with these simulations over the full wavelength range.

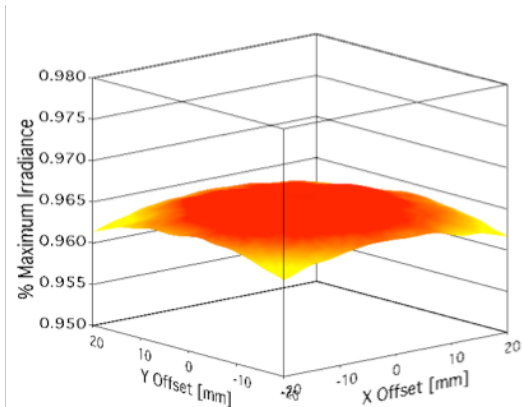

Figure 5. ZEMAX simulated focal plane irradiance of four point sources in the pupil plane.

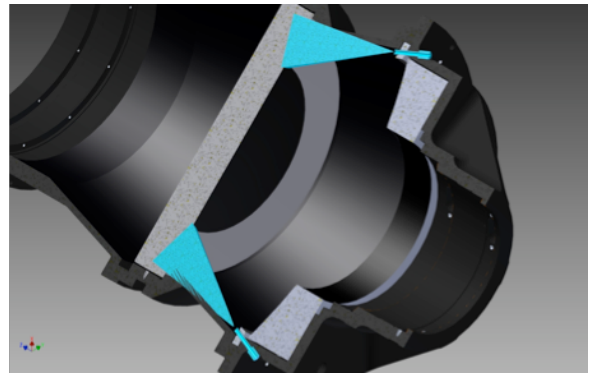

Figure 6. Cut-away drawing showing fiber optic positioning within the LSST Beam Simulator.

## 4. MECHANICAL SYSTEM

The optics is held in position with the mechanical assembly shown below in a cut-away drawing (Figure 7). A re-alignment mechanism is integrated with the optical assembly, and baffles are included. The optical assembly is mounted on an 8-foot optical table together with the scattering sphere, xyz positioner, dewar, electronics, and monochromator. The dewar+CCD may be moved in xyz by a programmable precision stage with additional azimuth angle adjustment. To enable efficient testing, we have two identical dewars. An important adjunct to the f/1.2 optical system is a class 1000 clean room with class 100 area for opening a dewar for installation of a new CCD for test.

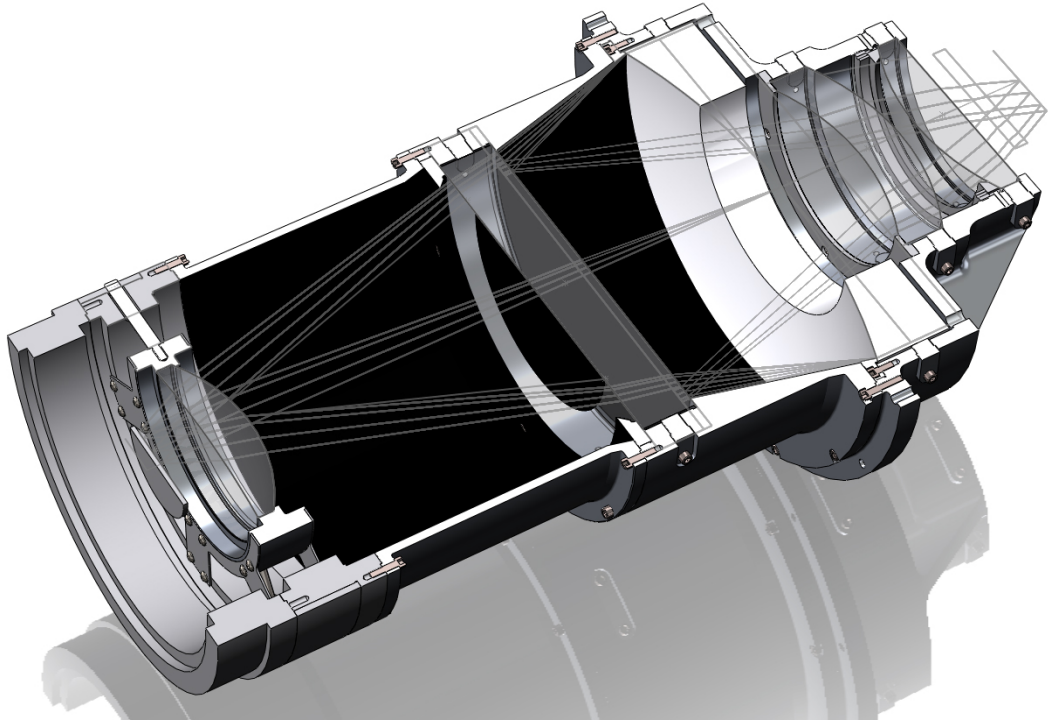

Figure 7. Cutaway of the f/1.2 system optical assembly showing three ray bundles. The central bundle is blocked by the central obscuration stop.

## 5. CCD TESTING

In order to study science capability, realistic sky scenes and spectral distributions are projected onto the full CCD focal plane. Since the LSST CCDs cover 350-1100 nm, the simulator covers this full range. Thick CCDs exhibit spatially dependent charge transport variations causing a mapping of input to output pixels in the 3-D device. The resulting astrometric and PSF shape effects are studied as a function of applied voltages, clock timings, light intensity and wavelength, background level, and temperature. Wavelengths of 0.7-0.9 micron are particularly important for precision weak gravitational lens measurements and require precision control of detector PSF shape systematics.

The wide-field f/1.2 reimaging system delivers an optical beam identical to that of the LSST camera with its central obscuration, and with sub pixel 5 micron FWHM PSF. The illumination, voltages, and camera position are controlled via LabView or Python. A variety of scenes can be inserted in the input, illuminated by a large scattering sphere. The fully automated system simulates LSST operation, including galaxies and stars superimposed on a spatially diffuse night sky emission with its complex spectral features.

The CCD can be positioned in the fast beam to better than 2 micron precision via 3-axis steppers. The system enables xy dithered imaging, as planned for LSST science operations[3]. A series of exposures are taken with the telescope moved in between. This automated "observing" mode of the re-imaging system provides the data necessary to assess the effectiveness of the software pipeline to correct for astrometric and photometric systematics.

The f/1.2 beam setup (Figure 8) allows a number of unique tests of the prototype science CCDs from vendors. These tests can be divided into several broad classes, depending on different aspects of the CCD response to realistic spectral and spatial illumination, exposure time, clocking rate, and opto-electronic behavior. In addition to the main imaging role for the CCDs, wavefront curvature detector tests and guider detector tests may also be carried out. The system has been used to fully characterize prototype CCDs. The output of the scattering sphere feeds through an optical bandpass filter and the source mask. Bandpass filters are provided for each of the LSST bands. These tests each utilize one or more mask/filter combinations:

- Filter only, no mask.  Used for flat-field direct illumination.
- Filter plus single pinhole mask (2-30 micron dia.).  To isolate effects near one location on the CCD.
- Filter plus masks with square arrays of pinholes (2-50 micron dia).  Astrometric residuals.
- Filter plus masks with random pinholes (2-50 micron dia).  Correlation tests.
- Filter plus masks with 1-D sine wave opacity modulation (various frequencies).  Astrometric residuals.
- Filter plus mask with artificial stars and galaxies.  Dithered observing tests of PSF-to-shear systematics.

Some of these tests use sky backgrounds of various levels and with realistic spectral features. Finally, for detailed studies vs wavelength, nearly monochromatic light from a scanning monochromator may be injected into the scattering sphere, in lieu of the bandpass filter at the re-imager input. Below we discuss some tests in more detail.

*Delivered PSF.*  The response of the device in an f/1.2 beam will vary with wavelength due to optical beam defocus in the silicon depletion volume, charge spreading, and CTE.  Charge collection and CTE will vary with back bias and clock rail voltages. It is possible to have temperature effects as well.  Spatial dependence of PSF over the chip is studied by moving the dewar in x and y with the programmable precision stage.  Sensitivity of delivered PSF to optical focal plane and device misalignment as a function of wavelength is also examined. Pixel-pixel correlation, including noise correlation, is studied.  For some of these tests, a source image consisting of 2-5 micron holes is used. Further tests are performed using realistic delivered optical PSFs (due to expected seeing over a range of 0.4 – 0.7 arcsec, depending on wavelength) utilizing a source mask consisting of 20-35 micron FWHM Gaussian "star" source mask images. A partial CCD image is shown in Figure 9.  Due to fringing fields, weak lensing shear systematic errors are introduced by non-square effective pixels near the edge of the CCD. This effect is studied by stepping the illuminating PSF array around in the edge area. Finally, the device photometric behavior (especially at 1 micron) as a function of defocus is studied.

*Delivered images.*  Imaging behavior of detectors depends on details of beam illumination, and device performance.  A source mask with a realistic density of stars and galaxies is used. The spectral energy distribution of these objects is adjustable, and a true night sky background spectrum is used.  This tests the imaging performance of the CCDs in realistic conditions in the LSST f/1.2 beam.  These tests can be done for different back bias, clock levels, and readout rates examining the S/N for galaxies and stars, in order to validate the LSST camera voltage choices.  Shift-and-stare imaging is done using the x-y motion of the dewar relative to the beam, and this data is jointly analyzed to search for low level systematics in the CCDs.  Finally, the dewar may be misaligned slightly relative to the optical axis in order to introduce a low level of image shear across the CCD, and this can be examined for E-mode and B-mode components as a test of the weak lens performance.

*Astrometric residuals.*  Using a dense grid of "stars" as shown in Figure 9 it is possible to map the astrometric error over the CCD.  This is examined on different spatial scales using "stars" of different diameter and dithers of different amplitude.  The method is as follows:   since the holes in the mask do not move relative to one another, they jointly define a coordinate system.  At each dither position this coordinate system can be determined by measuring the centroids of the holes [via weighted intensity first moments].   The dither offset vector may then be calculated to high precision from the grid coordinate offset in a dithered image.  Then for each of the "stars" the difference between its measured centroid and the expected centroid from the master dither offset vector is calculated, yielding the astrometric error for that star. This is repeated for each of the thousands of stars in the mask, generating an astrometric error map. To regularize this map, the astrometric residuals may be fit to a 3-D model of the CCD charge transport, if desired.

*Shape transfer function (STF) to shear residuals*  All of the above effects, in addition to charge transport and collection effects, influence the STF causing a mapping of PSF and galaxy shapes between that delivered by the photons at the focal surface and that delivered on the CCD pixelated output. We measure the STF with dithered arrays of stars. The output image shear is a convolution of the input image (PSF or galaxy) with the STF.  This STF mapping will be incorporated into the image processing pipeline as a WCS re-mapping. Finally, we plan to simulate LSST operations by "observing" fields of stars and galaxies. With this simulated observing data we can estimate the weak lens *shear residuals* after pipeline correction for the STF on realistic densities of star and galaxy images.

*Fringing.* At very long wavelengths thick fully depleted CCDs may exhibit slight fringing due to narrow night sky emission lines. The realistic sky spectrum generated by the spectral engine enables tests of the fringing in these devices in the f/1.2 LSST beam in a way that cannot be done with tests on other slower beam telescopes or in the lab. Subtle features in CCD response in actual cameras is common, and we plan a series of tests of these new CCDs in a realistic beam and spectral illumination. This is particularly true of the 1 micron band.

*Wavefront sensor (WFS) tests.* In addition to testing the science CCDs, the wavefront sensing (WFS) assembly (including the case where a defocused science 4K CCD is used for WFS) and algorithm can be tested. The wide field with hundreds of stars will enable lab tests of the curvature tomographic inversion algorithm and pipeline. This can be done even before construction using either out-of-focus science CCDs or split CCDs. Controlled aberrations can be introduced and output error signals compared. This can be repeated under a variety of conditions such as vignetting, WFS tilt, star crowding, stellar density, and S/N ratio. This will occur early during construction, enabling assessment and refinement of the algorithm and optimization of performance under various observing conditions.

*Guider tests.* The baseline plan for the guide CCDs is to utilize a full 4K science CCD in a "region of interest" mode where fast clocking in two directions is used. This has never been done before, and the analysis of its performance for our segmented thick CCDs assumes no asymmetric systematic such as CTE. This test stand and electronics will allow us to check the performance of these 4K CCDs under this novel application. The performance is expected to depend on the night sky spectral illumination as well as clocking. The test stand permits the appropriate spectral illumination.

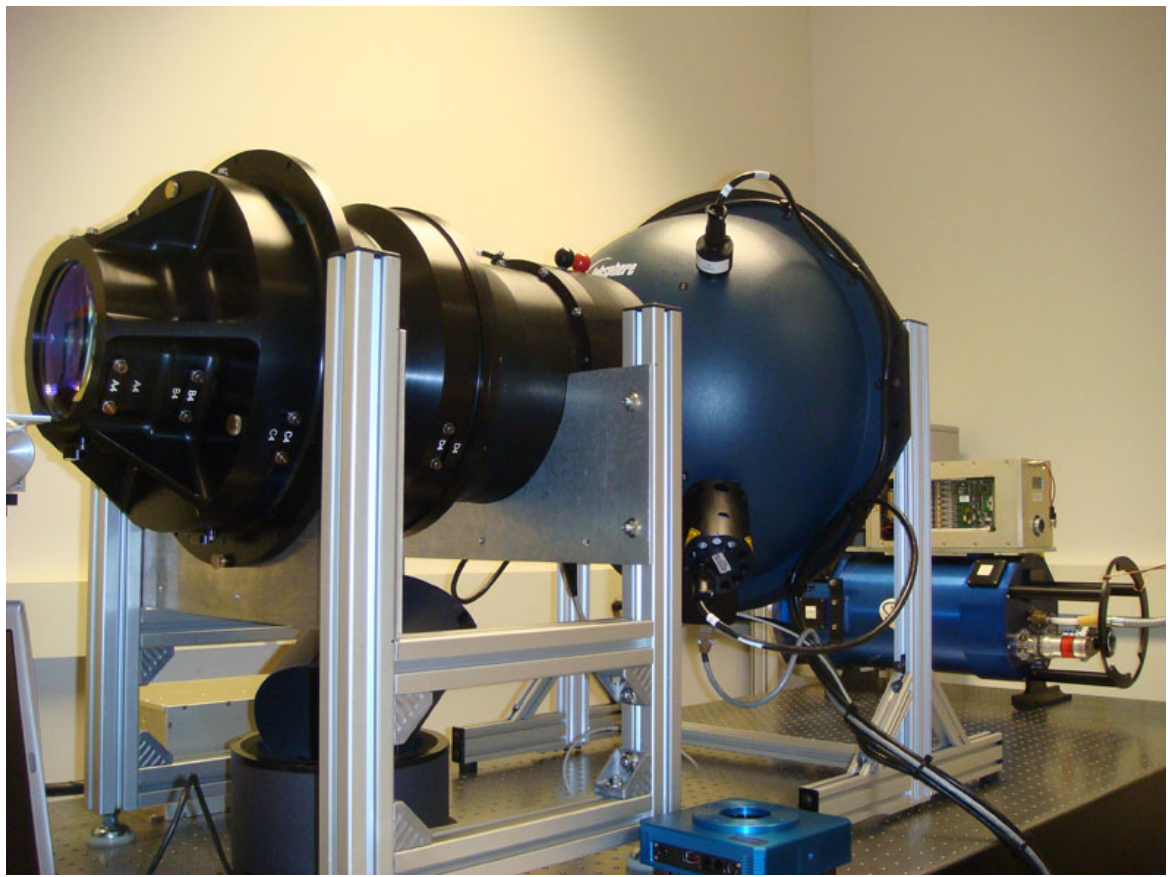

Figure 8. The f/1.2 system optical assembly and scattering sphere, with the camera dewar and electronics in the background. The 145 mm diameter exit optic can be seen on the left end of the reimager. The precision positioning stage is off the picture to the left.

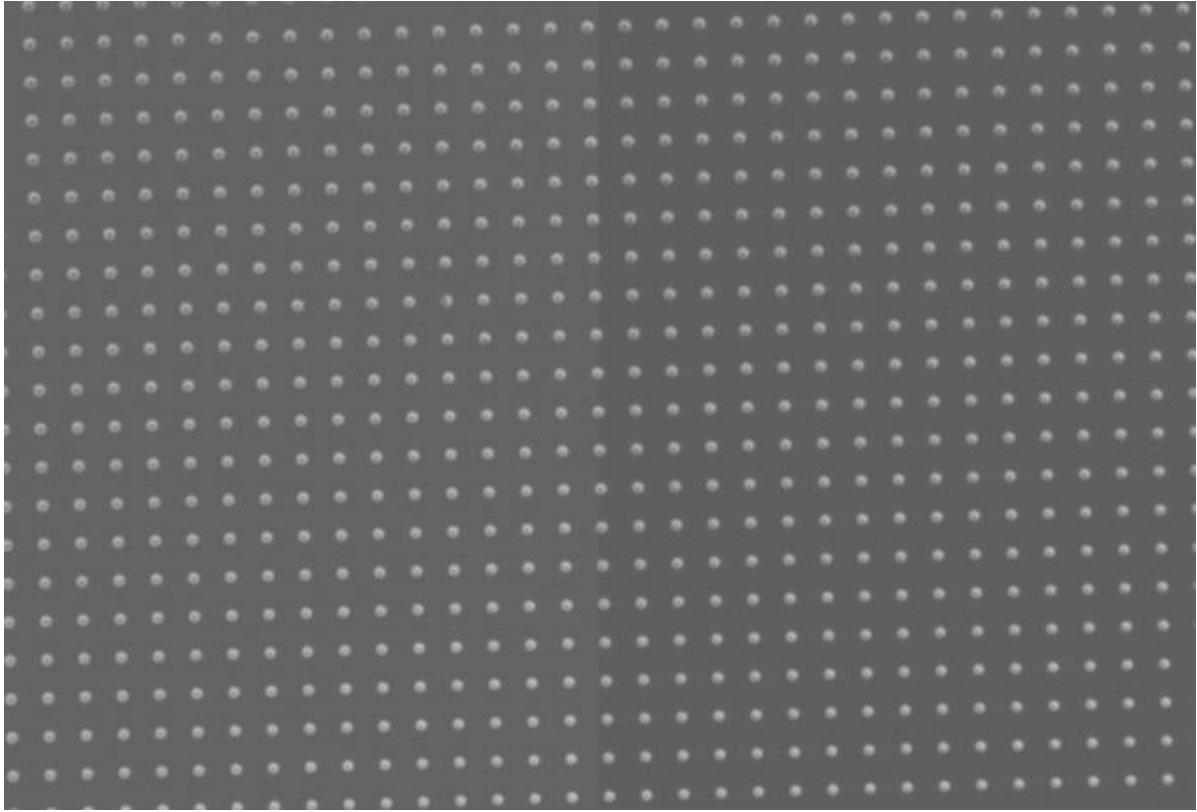

Figure 9. In one test protocol the 4Kx4K CCD is illuminated with 40,000 artificial stars: a grid of 30 micron diameter holes spaced 200 microns. A small portion of the CCD is shown here in a raw image with no bias (zero level) correction. The displayed region straddles two of the 16 sub-CCD segments. The mask is tilted slightly, relative to the CCD columns, to increase the spatial sampling on small scales. Thousands of xy dithered exposures are taken as a function of light level and voltages, enabling precision reconstruction of astrometric and photometric residuals and characterization of the PSF shape transfer function as a function of position on the CCD.

## ACKNOWLEDGEMENTS

We thank Leo Alcorn, Ian Crane, Perry Gee, Chris Stubbs, Peter Takacs, and John Warren for their help. We acknowledge support from NSF AST ATI grant 0441069, the W. M. Keck Foundation, DOE grant DE-SC0009999, Wayne Rosing and Dorothy Largay, Eric Schmidt, and the National Optical Astronomy Observatory.